\documentclass[aps,twocolumn,prl,amsmath,amssymb,floatfix,superscriptaddress]{revtex4-1}

\usepackage{natbib}
\usepackage[latin9]{inputenc}
\usepackage{amsmath}
\usepackage{amssymb}
\usepackage{graphicx}
\usepackage{esint}
\usepackage{braket}
\usepackage{multirow}
\usepackage{dcolumn}
\usepackage{color}

\makeatletter
\makeatletter
\newcommand{\unit}[1]{\ensuremath{\,\mathrm{#1}}}

\newcommand{\MHz}{\unit{MHz}}
\newcommand{\KHz}{\unit{kHz}}
\newcommand{\GHz}{\unit{GHz}}

\newcommand{\us}{\unit{\mu s}}

\newcommand{\Yb}{\ensuremath{^{171}\mathrm{Yb}^+~}}
\newcommand{\avg}[1]{\ensuremath{\left\langle#1\right\rangle}}

\begin{document}
\title{Quantum simulation of the quantum Rabi model in a trapped ion}

\author{Dingshun Lv$^{1}$, Shuoming An$^{1}$, Zhenyu Liu$^{1}$, Jing-Ning Zhang$^1$, \\ Julen S. Pedernales$^{2,3}$, Lucas Lamata$^{2}$, Enrique Solano$^{2,4}$, \& Kihwan Kim$^{1}$}

\affiliation{$^{1}$Center for Quantum Information, Institute for Interdisciplinary Information Sciences, Tsinghua University, Beijing 100084, P. R. China  \\ $^{2}$Department of Physical Chemistry, University of the Basque Country UPV/EHU, Apartado 644, 48080 Bilbao, Spain \\ $^3$ Institute for Theoretical Physics and IQST, Albert-Einstein-Allee 11, Universit\"at Ulm, D-89069 Ulm, Germany \\$^{4}$IKERBASQUE, Basque Foundation for Science, Maria Diaz de Haro 3, 48013 Bilbao, Spain
}

\begin{abstract}
The quantum Rabi model, involving a two-level system and a bosonic field mode, is arguably the simplest and most fundamental model describing quantum light-matter interactions. Historically, due to the restricted parameter regimes of natural light-matter processes, the richness of this model has been elusive in the lab. Here, we experimentally realize a quantum simulation of the quantum Rabi model in a single trapped ion, where the coupling strength between the simulated light mode and atom can be tuned at will. The versatility of the demonstrated quantum simulator enables us to experimentally explore the quantum Rabi model in detail, including a wide range of otherwise unaccessible phenomena, as those happening in the ultrastrong and deep strong coupling regimes. In this sense, we are able to adiabatically generate the ground state of the quantum Rabi model in the deep strong coupling regime, where we are able to detect the nontrivial entanglement between the bosonic field mode and the two-level system. Moreover, we observe the breakdown of the rotating-wave approximation when the coupling strength is increased, and the generation of phonon wave packets that bounce back and forth when the coupling reaches the deep strong coupling regime. Finally, we also measure the energy spectrum of the quantum Rabi model in the ultrastrong coupling regime.
\end{abstract}

\maketitle

\section*{Introduction}
The interaction between light and matter is one of the most fundamental and ubiquitous physical processes. The semi-classical Rabi model was proposed in 1936 to analyze the effect of a varying, weak magnetic field on an oriented atom possessing nuclear spin \cite{rabi1936process}. It describes the dipolar interaction between a classical monochromatic field and a two-level system, successfully explaining the challenging experimental data in Ref.~\cite{frisch1933einstellung}. When the field is promoted to a quantum description, resulting in the simplest fully-quantum model of light-matter interaction, it is called the quantum Rabi model (QRM). Typically, the coupling strength in a light-matter system is much lower than the field frequency. In this scenario, the QRM can be simplified to the Jaynes-Cummings model (JCM) under the rotating-wave approximation (RWA)~\cite{Jaynes63}. The JCM is an analytically solvable model that has been studied in cavity quantum electrodynamics(CQED)~\cite{miller2005trapped,walther2006cavity,Haroche}, atomic physics~\cite{Ritsch2013Cold}, quantum dots~\cite{Hanson2006Spins}, circuit quantum electrodynamics (cQED)~\cite{wallraff2004strong,devoret2013superconducting} and trapped ions~\cite{meekhof1996generation,Leibfried03,Haffner08,lv2017reconstruction}, among other quantum platforms. Recent experimental achievements have shown the accessibility to the ultrastrong coupling (USC) regime\cite{niemczyk2010circuit,Braum2016Analog} or even to the deep strong coupling (DSC) regime~\cite{forn2017ultrastrong,yoshihara2017superconducting}, where the coupling strength is comparable to or larger than the mode frequency. In these strong-coupling regimes, the RWA breaks down, rendering the JCM as a restricted description of the system, and requiring the use of the full QRM to correctly describe the emerging physical phenomena~\cite{Rossatto2016Spectral}. It is noteworthy to mention that, in such regimes, exotic dynamical properties of light-matter interaction~\cite{casanova2010deep} and potential applications to quantum information technologies \cite{romero2012ultrafast} have been predicted and proposed.

The Hamiltonian associated with the QRM can be expressed as ($\hbar=1$)

\begin{eqnarray}{\label{JCnAJC}}
\hat{H}_{\rm QRM} =
\frac{\omega_{0}}{2}\hat{\sigma}_{z}+\omega_{m}\hat{a}^{\dag}\hat{a}+ g\hat{\sigma}_{x}(\hat{a}+\hat{a}^{\dag}),
\label{QRM}
\end{eqnarray}
where $\hat{a}^{\dag}$($\hat{a}$) is the creation(annihilation) operator related to a bosonic mode with frequency $\omega_{m}$, $\hat{\sigma}_{z,x}$ are the Pauli operators acting on a two-level system with energy splitting $\omega_{\rm{0}}$, and $g$ is the coupling strength. Three major coupling regimes are defined depending on the ratio between the coupling strength $g$ and the field mode frequency $\omega_m$, namely, the Jaynes-Cummings regime, with $\frac{g}{\omega_m} \ll 0.1$ and where the RWA is valid, the USC regime with $0.1\lesssim\frac{g}{\omega_m}$ and where the RWA is not a valid approximation, and the DSC regime with $1 \lesssim\frac{g}{\omega_m}$. Regardless of the apparent simplicity of its succint Hamiltonian form, an analytical solution of the QRM has only recently been found~\cite{braak2011integrability}. Nowadays, experimental efforts have been made to reach the DSC regime \cite{forn2017ultrastrong,yoshihara2017superconducting}, which allows for the study of exotic physics such as the bouncing back and forth of the photon number wave packet along the parity chains, and the entangled nature of the ground state of the system~\cite{casanova2010deep}. The phenomenon of photon number wave packets bouncing back and forth has been observed in a classical simulator of a photonic waveguide system \cite{crespi2012photonic} and in an analogue and a digital quantum simulation with cQED systems \cite{Braum2016Analog,langford2016experimentally}. However, the study of the ground state in DSC regime is still an open challenge.

In this work, we report the analog quantum simulation of the quantum Rabi model with a single trapped ion for all relevant coupling regimes. Among other results, we generate and observe the ground state of the QRM in the DSC regime in a trapped-ion quantum simulator for the first time. We demonstrate the full controllability and tunability of the QRM in a single trapped-ion system as proposed in Ref.~\cite{pedernales2015quantum}, which enable us to generate the exotic ground state in the DSC regime by the adiabatic transfer from the simple ground state of the JCM. Moreover, we apply the capability of the ground state preparation to experimentally measure the energy spectrum of the QRM Hamiltonian (\ref{JCnAJC}).

\section*{Trapped-ion system}
In our experiment, a radio-frequency Paul trap is used to spatially confine an $\Yb$ ion that is then cooled down to its motional ground state by standard sideband cooling techniques~\cite{Leibfried03} after Doppler cooling. In the low-energy regime, the motion of the ion can be well approximated to that of a harmonic oscillator, and two energy levels of the hyperfine manifold of its electronic ground state can be used as a qubit. In particular, we encode the two-level system in the levels $\ket{F=1,m_{F}=0}\equiv\ket{\uparrow}$ and $\ket{F=0,m_{F}=0}\equiv\ket{\downarrow}$ of the $S_{1/2}$ hyperfine manifold, which have a transition frequency $\omega_{\rm HF} = (2 \pi) ~ 12.642812$ \GHz. We employ a radial vibrational mode of frequency $\omega_{\rm X}$ = (2$\pi$) 2.498 $\MHz$ as the bosonic degree of freedom of our simulator. The uncoupled Hamiltonian describing such a system is given by $\hat{H_{0}}= \frac{\omega_{\rm{HF}}}{2}\hat{\sigma}_z+\omega_X\hat{a}^{\dag}\hat{a}$. When a pair of counterpropagating Raman laser beams is driven onto the ion, the general ion-laser interaction is described by
\begin{eqnarray}
\hat{H}_{\rm ion-laser} = \Omega_{\rm l} \cos\left( \Delta k_{\rm l} \hat x - \omega_{\rm l} t + \phi_{\rm l} \right) \hat{\sigma}_{\rm x}.
\label{ionlaserHam}
\end{eqnarray}
Here, $\Omega_{\rm l}$ is the Rabi coupling strength proportional to the product of both laser field amplitudes, $\Delta k_{\rm l}$ is the net wave vector component of the Raman laser beams on the direction of the motion of the ion, $\hat x = x_{0} (\hat{a}+\hat{a}^{\dagger})$ is the position operator of the ion, with ground-state wave-packet width $x_0=\sqrt{\hbar/ 2M\omega_{\rm X}}$, where $M$ is the mass of the \Yb ion, and $\omega_{\rm l}$ and $\phi_{\rm l}$ are the differences of frequencies and phases of the Raman laser beams, respectively \cite{lv2017reconstruction}.

Under suitable optical and vibronic RWAs, and also in the Lamb-Dicke regime, the ion-laser interaction can be turned into a(n) (anti-)Jaynes-Cummings interaction by tuning the laser frequency close to the red(blue)-sideband as ${\omega_{\rm r}=\omega_{\rm HF} - \omega_{\rm X} - \delta_{\rm r}}$ (${\omega_{\rm b}=\omega_{\rm HF} + \omega_{\rm X}-\delta_{\rm b}}$), with a small detuning $\delta_{\rm r(b)} \ll \omega_{\rm X} $ in the most general case. Red-sideband and blue-sideband interactions are described in the interaction picture by the Hamiltonians~\cite{lv2017reconstruction,an2014experimental}
\begin{eqnarray}
\hat{H}_{\rm red} \left(\rm t \right)&=& \frac{\eta\Omega_{\rm r}}{2}\left(\hat{a}\hat{\sigma}_{+} \mbox{e}^{i \delta_r t} +\hat{a}^{\dag}\hat{\sigma}_{-}\mbox{e}^{-i \delta_r t}\right), \nonumber \\ \hat{H}_{\rm blue} \left(\rm t \right) &=& \frac{\eta\Omega_{\rm b}}{2}\left(\hat{a}^{\dag}\hat{\sigma}_{+} \mbox{e}^{i \delta_b t}+\hat{a}\hat{\sigma}_{-} \mbox{e}^{-i \delta_b t}\right) .
\end{eqnarray}
Here, $\eta=\Delta k_{\rm l} x_0 $ is the Lamb-Dicke parameter, and $\hat{\sigma}_{+} \left(\hat{\sigma}_{-}\right)=\ket{\uparrow}\bra{\downarrow} (\ket{\downarrow}\bra{\uparrow})$ is the spin-raising (lowering) operator.

When both red and blue sideband interactions are simultaneously applied with equal strength, such that $\Omega=\Omega_{\rm r}=\Omega_{\rm b}$, one can write the total Hamiltonian in the interaction picture as
\begin{eqnarray}
\hat{H}_{\rm br} \left(\rm t \right)&=& \frac{\eta\Omega}{2} \hat{\sigma}_{+}\left(\hat{a} \mbox{e}^{i \delta_r t} +\hat{a}^{\dag}\mbox{e}^{i \delta_b t}\right)+\rm{H.c.}
\label{QRMExp}
\end{eqnarray}

Indeed, Eq.~(\ref{QRMExp}) corresponds to the Hamiltonian of the QRM in Eq.~(\ref{QRM}) considered in the interaction picture with respect to the uncoupled Hamiltonian $\hat{H}^{'}_0= \frac{\delta_{b} + \delta_r}{4}\hat{\sigma}_{z} + \frac{\delta_b - \delta_r }{2}\hat{a}^{\dag}\hat{a}$. Therefore, if we undo the interaction picture transformation, we have
\begin{eqnarray}
\hat{H}_{\rm eff} &=&\frac{(\delta_{\rm{b}} +\delta_{\rm{r}})}{4}\hat{\sigma}_{\rm z} +\frac{(\delta_{\rm{b}}-\delta_{\rm{r}})}{2}\hat{a}^{\dagger}\hat{a}\nonumber \\ &+& \frac{\eta\Omega}{2} (\hat{\sigma}_{\rm +}+ \hat{\sigma}_{\rm -})(\hat{a}+\hat{a}^{\dagger}),
\label{Heff}
\end{eqnarray}
where the parameters of the simulated QRM can be associated with the experimental ones as ${\omega_0 =\frac{\delta_{\rm{b}} +\delta_{\rm{r}}}{2}}$,  ${\omega_m =\frac{\delta_{\rm{b}}-\delta_{\rm{r}}}{2}}$ and $g=\frac{\eta\Omega}{2}$. Thus, such an experimental setup serves as a quantum simulator of the QRM, where one can simulate a wide range of coupling regimes by suitably tuning the laser intensities and detunings to match the desired ratio $\frac{g}{\omega_m}$. It is important to point out that the observables of interest $\{ \hat{a}^{\dag}\hat{a},\hat{\sigma}_{z},\ket{n}\bra{n} \}$, commute with all the adopted interaction-picture transformations, which are always with respect to a Hamiltonian of the form $\alpha \hat{a}^{\dag}\hat{a} +\beta \hat{\sigma}_{z} $, such that their expectation values will remain unaltered in the laboratory reference frame~\cite{pedernales2015quantum}.

\section*{Coupling regimes and breakdown of the RWA}

\begin{figure*}[ht]
\includegraphics[width=1.0\textwidth]{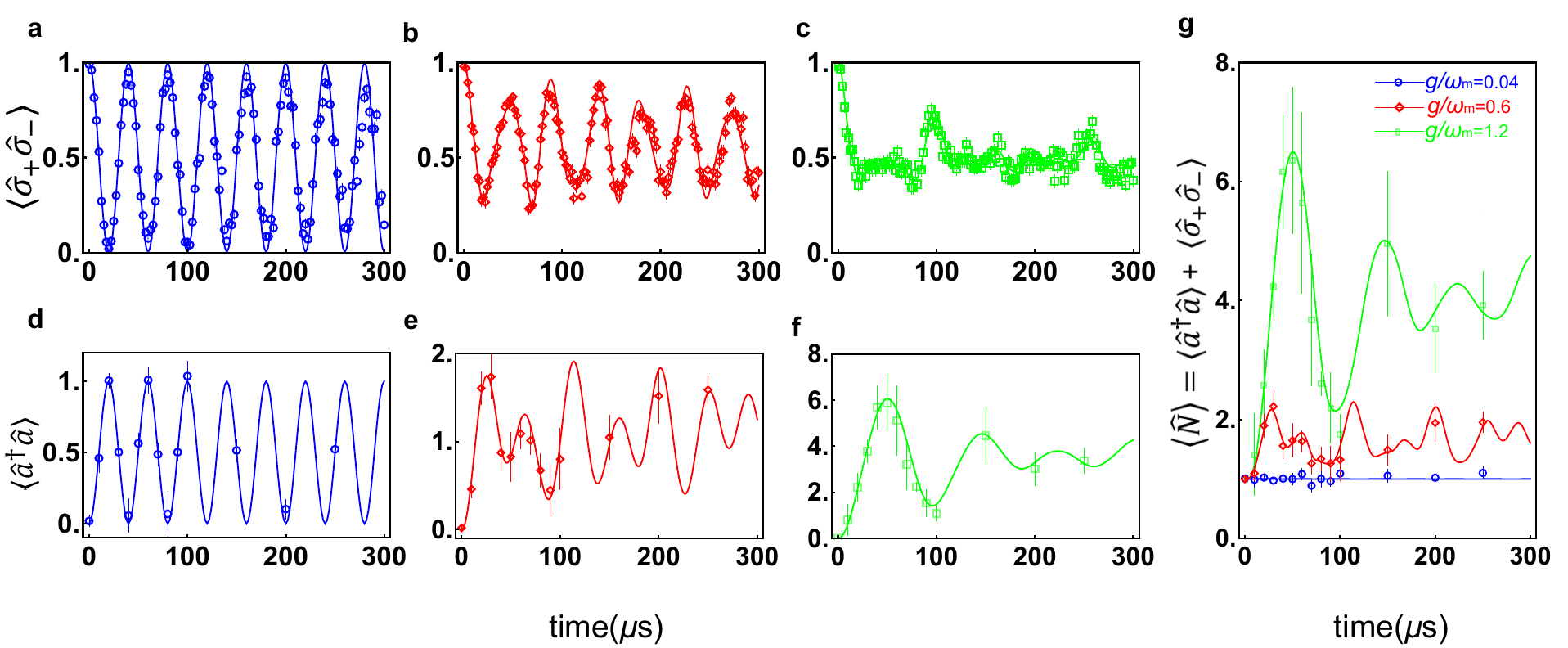}
\caption{{\bf Spin and phonon dynamics under the QRM for different coupling regimes.} (a, b, c) correspond to the population of the excited state of the two-level system for the coupling ratios $g/\omega_m=0.04$, 0.6, and 1.2, respectively. (d, e, f) correspond to the evolution of the average number of phonons for the same coupling ratios. Finally,  (g) shows the evolution of the total number of excitations for the three cases considered above. In all panels, theoretical predictions are plotted with continuous lines, while dots and their associated error bars represent the experimental data. \label{fig:DCS}}
\end{figure*}

For the experiment, we fix the coupling strength to $g=(2\pi)12.5 \KHz$, and the detuning of the red sideband to $\delta_{\rm{r}} =0$, leaving $\delta_{\rm{b}}$ as a tunable parameter. In this manner we will be simulating a resonant QRM where the ratio $\frac{g}{\omega_m}$ will be determined by the selected detuning $\delta_{\rm{b}}$. We experimentally explore three paradigmatic coupling regimes, namely the Jaynes-Cummings, the USC and the DSC regimes, accordingly selecting the values of the detuning for the blue sideband as ${\delta_{\rm{b}}=2\omega_{m}=(2\pi)\{ 625,\ 83.4, \  41.6\} \KHz}$, which correspond to the ratios $g/\omega_m = \{ 0.04,\ 0.6,\ 1.2 \}$, respectively.

The experiment is carried out as follows. First, we perform standard Doppler and sideband cooling, which prepares the system in the state $\ket{\downarrow,n=0}$ \cite{sidebandcooling}, and then, we transfer the system to the initial state $\ket{\uparrow,n=0}$ by applying a carrier $\pi$ pulse. After that, we turn on the red-sideband and blue-sideband transitions, with suitably chosen intensities and detunings, to implement the QRM Hamiltonian in the desired regime. We observe the dynamics of the QRM by measuring the average excitations of the spin $\avg{\hat{\sigma}_{+}\hat{\sigma}_{-}}$ and the phononic degrees of freedom $\avg{\hat{a}^{\dagger}\hat{a}}$ at specific evolution times $t$ (see Methods).

In Fig.~\ref{fig:DCS}(a) and (d), the measurements for the simulation of the Jaynes-Cummings regime are plotted. Rabi oscillations, with a complete collapse and posterior revival of the excitation probability of the two-level system are clearly observed. In the same manner, the average number of phonons in the bosonic mode oscillates between 0 and 1, consistent with the notion that the wavefunction of the system should live in the space spanned by the corresponding JCM doublet $\{ | 0, \uparrow \rangle, | 1, \downarrow \rangle \}$, as expected for such a regime. Figures \ref{fig:DCS}(b) and (e) show the evolution of the same initial state in the USC regime for the coupling ratio $g/\omega_m=0.6$. In this case, collapses and revivals of the excitation probability are not complete and the average number of phonons exceeds 1, indicating that the dynamics does not anymore happen exclusively in the JCM doublet. This departure from the JCM physics is associated with the breakdown of the RWA due to the large coupling ratio. In the DSC regime, plotted in Figs.~\ref{fig:DCS}(c) and (f) for the coupling ratio $g/\omega_m=1.2$, the effects of the RWA breakdown are even clearer, where not even oscillations can be identified and where the average number of phonons grows above 6 for the plotted example. We also show in Fig.~\ref{fig:DCS}(g) the evolution of the total excitation number $\langle\hat{N}\rangle=\langle\ket \uparrow\bra \uparrow\rangle +\langle\hat{n}\rangle$, which is a conserved quantity when the RWA is valid, but has a dynamical behavior as soon as the RWA breaks down.

\section*{DSC regime and phonon wavepackets}

\begin{figure*}[ht]
\includegraphics[width=1.0\textwidth]{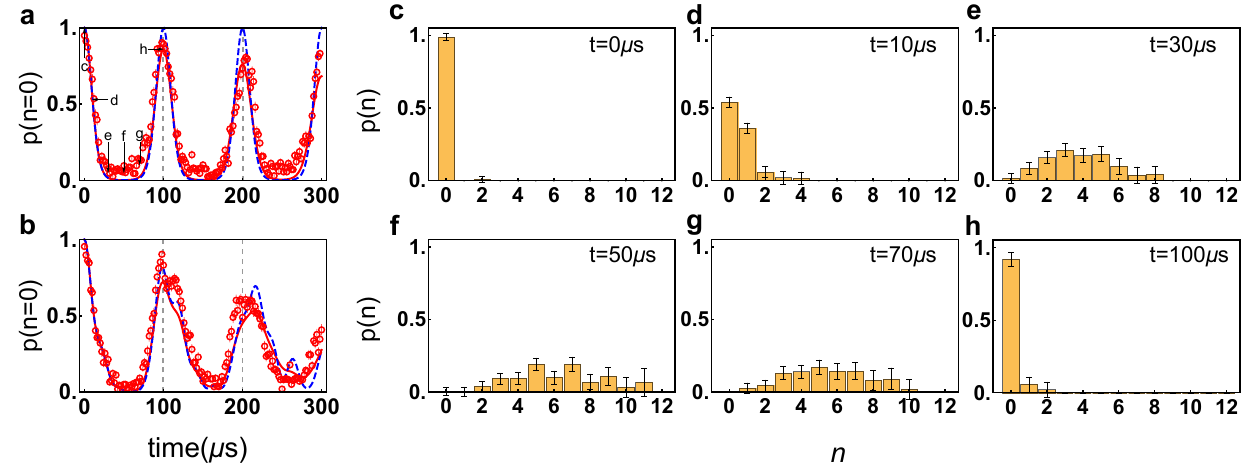}
\caption{{\bf Phonon-number wave packets bouncing back and forth in the DSC regime.} In ({\bf a}) and ({\bf b}) we plot the population of state $\ket{n=0}$, after tracing out the spin, as the system evolves under the QRM in the DSC regime. In particular, ({\bf a}) shows the spin-degenerate case, $\omega_{\rm{0}} =0$, and ({\bf b}) the non-degenerate case, with $\omega_{\rm{0}} = 0.8 g$. For both cases, the coupling ratio is fixed to $\frac{g}{\omega_{\rm{m} }}=1.25$. Dashed and solid lines represent theoretical calculations with and without decoherence of the motional mode, respectively. The vertical dashed lines indicate the motional revival times $k \frac{2\pi}{\omega_{\rm{m} }}$, where $k = 1,2,... $. The data points with error bars correspond to the experimental results. We obtain the zero-phonon population following the method in Ref.~\cite{lv2017reconstruction} after tracing out the spin. ({\bf c-h}) show the phonon number distribution sampled at several instants during one period $T=\frac{2\pi}{\omega_m}=100\us$ of the QRM Hamiltonian for the spin degenerate case. The phonon distribution is obtained by fitting the standard blue-sideband signals after tracing out the spin (see Methods). At the first revival time, the phonon state is back to the initial state as predicated by the QRM. The imperfections can be attributed to decoherence of the motional degrees of freedom.
\label{fig:PhononBouncAndBack}}
\end{figure*}

We focus now on the DSC regime and explore two scenarios, namely the case of the degenerate QRM with $\omega_{0}= 0$ and the non-degenerate case with $\omega_{0} \neq 0$. For this experiment, we fixed the coupling strength to $g=(2\pi)12.5 \KHz$, and vary $\delta_{\rm{r}} $ and  $\delta_{\rm{b}} $, while keeping always a ratio $g/\omega_m=1.25$. For the degenerate case, we use detunings  $\delta_{\rm{r}}=\delta_{\rm{b}}=(2\pi)10 \KHz$, while for the non-degenerate case we use $\delta_{\rm{r}}=0$ and $\delta_{\rm{b}}=(2\pi)20 \KHz$, which corresponds to $ \omega_{0}=0.8g$. For the initial state, we choose the ground state of the JC model $\ket{\downarrow,0}$, which should have no dynamical properties when the coupling strength $g/\omega_m$ is small enough. In Fig.~\ref{fig:PhononBouncAndBack}, we show that the situation is different when considering the DSC regime. First of all, in panels (a) and (b), we show the evolution of the population of Fock state $\  | 0 \rangle$, after tracing out the spin degree of freedom. For the spin degenerate case, we can clearly observe that this population collapses to zero and that it is stabilised at zero except for every one period of the mode frequency, that is to say, except for times $k\frac{2\pi}{\omega_m }$, $k$ being an integer, where a full revival of the population is detected. On the other hand, the non-degenerate case shows a degradation of these revivals for long times, as it was predicted in Ref.~\cite{casanova2010deep}. Additionally, we sample several points during one period $T=\frac{2\pi}{\omega_m}=100 \mu s$ of the evolution of the degenerate case and measure its phonon distribution, as shown in panels~(c-h). We obtain the phonon number distributions by fitting the spin-excitation evolution under the blue-sideband transition with the function shown in the Methods section. At time zero, the population is concentrated on Fock state $| 0 \rangle$, and as time elapses higher Fock states are populated. The evolution resembles a wavepacket that travels along a chain of Fock states up to a maximum determined by $\sim 4(g/\omega_m)^2$ and then comes back to the initial states at one period of the mode frequency. This phenomenon was theoretically predicted in Ref.~\cite{casanova2010deep} as characteristic of the DSC regime, and it is referred to as the bouncing back and forth of phonon-number wave packets. We note that the $\omega_m =0$ case has already been studied in the literature for its application as a geometric phase gate, where the effects over the spin degrees of freedom are the main interest \cite{S2000Entanglement,Haljan2005Spin}.

\section*{Adiabatic ground-state preparation}

\begin{figure*}[ht]
\includegraphics[width=1.0\textwidth]{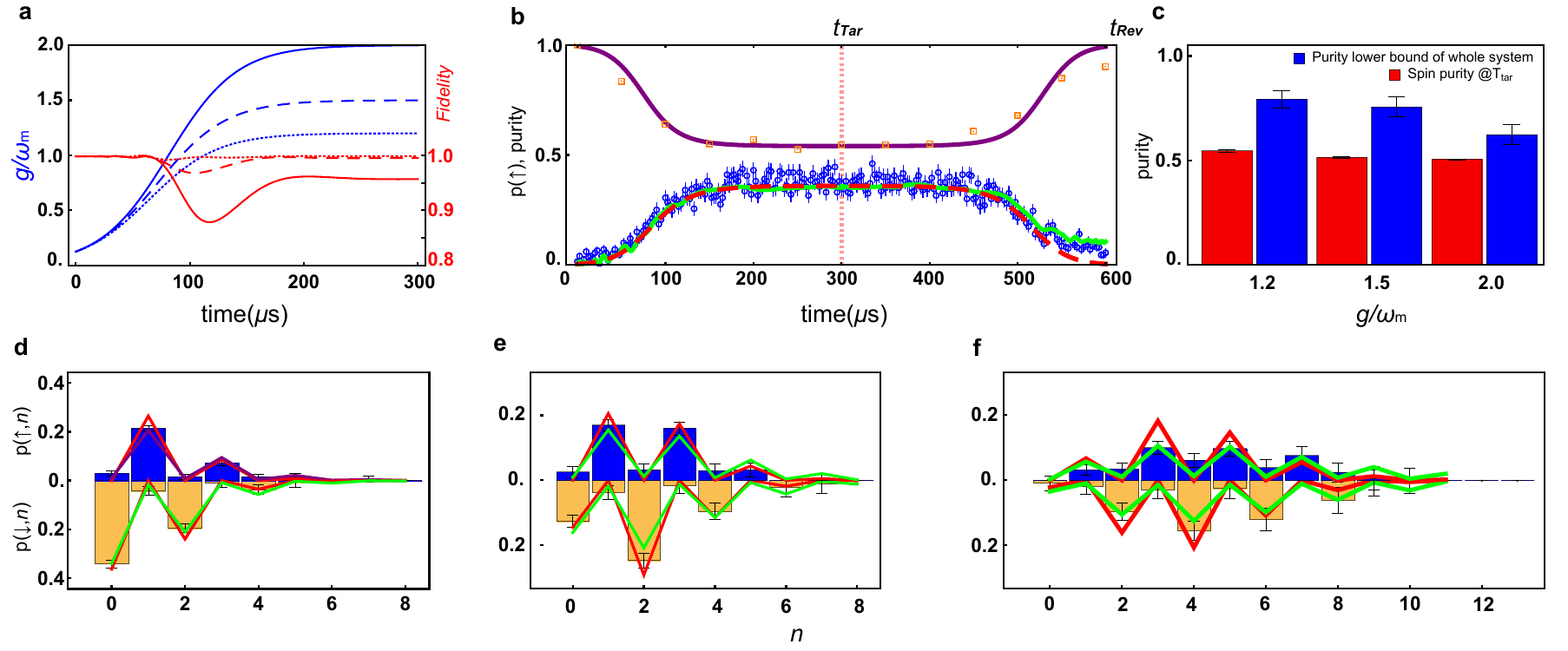}
\caption{{\bf Adiabatic ground state preparation of the QRM in the DSC regime.} In panel {\bf (a)}, we show the adiabatic scheme for the preparation of the ground state of the QRM at the DSC ratios $\frac{g}{\omega_{\rm{m}} }=1.2, 1.5,$ and $2.0$, as starting from the initial JC ratio $\frac{g}{\omega_{\rm m}}=0.125$. The fidelities between the instantaneous ground state and the numerically evolved state quantify the adiabaticity of our process. In panel {\bf (b)}, we show the evolution of the excitation probability and the purity of spin state during the adiabatic ground state preparation ($0\leq t \leq t_{\rm Tar}$) and during the reverse process ($t_{\rm Tar}\leq t \leq t_{\rm Rev}$). The plot corresponds to the preparation of the ground state at the ratio $\frac{g}{\omega_{\rm{m}} }=1.2 $. The red dashed line and the green solid line are obtained by direct diagonalization of the QRM Hamiltonian and numerical simulation of the adiabatic process, respectively, including heating and dephasing of the motional mode, which are expected experimental imperfections. The blue circles with error bars correspond to the experimental results. The purple line represents the numerically computed purity of the spin, defined as $\rm{Tr}(\rho_{\rm spin}^{2})$, where $\rho_{\rm spin}$ is the reduced density matrix of the spin after tracing out the motional degree of freedom. The orange squares are the corresponding experimental results, computed from the spin-tomography. Panels {\bf (d)} to {\bf (f)} show the phonon-number distributions correlated with $\ket{\downarrow}$ (lower panel) and $\ket{\uparrow}$ (upper panel), which are obtained by fitting the standard blue-sideband signals after the spin-projective measurement for $\frac{g}{\omega_{\rm{m} }}=1.2, 1.5,$ and 2.0, respectively. Finally, panel {\bf (f)} shows the purity of the spin and estimated lower bounds for the purity of the whole system, for each tested coupling ratio. The purity of the spin is obtained from the measured reduced density matrix, which was done by spin-tomography. The lower bounds of the purities of the whole system are estimated by measuring the probability of being in $\ket{\downarrow,0}$ after the reverse process at $t_{\rm Rev}$.
\label{fig:GS}}
\end{figure*}

As mentioned in the previous section, the ground state of the QRM in the Jaynes-Cummings regime $(g \ll \omega_m )$ is given by the state $\ket{\rm{\downarrow},0}$, while the ground state of the QRM in the DSC regime is a nontrivial state where spin and field are entangled, and which to the best of our knowledge has never been implemented in a physical quantum platform.

In our experiment, we generate the ground state of the QRM in the DSC regime by starting in the ground state of the Jaynes-Cummings regime, the state $\ket{\rm{\downarrow},0}$, and adiabatically increasing the coupling ratio $\frac{g}{\omega_m }$ towards the DSC regime. To achieve this, one could choose to either increase $g=\eta \Omega/2$, which can be done by rising the laser intensity, or decrease ${\omega_m=(\delta_r - \delta_b)/2}$. Because it is experimentally more feasible to manipulate the detuning of the Raman lasers than their power, we choose the latter in our experiment. For that, we fix coupling strength $g$ to be $(2\pi)12.5 \KHz$ and $\delta_{\rm{r}}=0 $, leaving $\delta_{\rm{b}}$ as the only tunable parameter, which we manipulate with an exponential time dependence of the form $\delta_{\rm{b}}(\rm t)= (\delta_{\rm{Max}} - \delta_{\rm{Tar}}) \mbox{e}^{-\frac{t}{\tau}} + \delta_{\rm Tar}  $. Here, we set $\delta_{\rm{Max}}=(2\pi) 0.2 \MHz$, while $\delta_{\rm{Tar}}$ is determined by the ratio $\frac{g}{\omega_m }$ we want to reach, and $\tau=\frac{t_{\rm Tar}}{10}$, $t_{\rm Tar}=300\mu s$ being the total duration of the adiabatic process. The adiabaticity of our scheme is guaranteed by the numerical computation of the fidelity between the instantaneous ground state of the Hamiltonian and the adiabatically evolved state, which is shown in Fig. \ref{fig:GS}a.

In panel (b) of Fig.~\ref{fig:GS}, we show the spin evolution during the adiabatic process for the time interval (0-$t_{\rm Tar}$). The plot corresponds to the case $g/\omega_{\rm{m}}=1.2$, with the cases for other ratios showing similar behavior. At time $t_{\rm Tar}$, the system is expected to be in the ground state of the QRM for the selected coupling regime. In panels (d) to (f), we plot the outcome of the phonon distributions correlated with the spin performed at this time for coupling ratios $g/\omega_m=1.2$, $1.5$ and $2.0$, respectively (see Methods). To verify the quantum coherence maintained within the preparation of the ground state, we reverse the adiabatic process in an attempt to recover the initial ground state~$\ket{\rm{\downarrow},0}$. In panel (b), we can observe how the spin returns to state~$\ket{\downarrow }$. As a complementary proof, we plot the purity of the spin state. To this aim, we trace out the phononic degrees of freedom and measure the density matrix $\rho$ associated with the spin degree of freedom \cite{Nielson2000Quantum}, from which we calculate the purity, defined as Tr($\rho^{2}$), during the whole process. The degradation of the purity during the preparation of the ground state of the QRM in the DSC regime confirms that the adiabatically prepared ground state is indeed an entangled state, and the subsequent revival of the purity when the adiabatic process is inverted proves that we are able to recover the initial state and therefore that the whole process preserves quantum coherence.

On the other hand, the expectation values of the parity operator $\Pi=\sigma_{z}\mbox{e}^{-i\pi\hat{a}^{\dagger}\hat{a}}$, for the states in panels (d) to (f) of Fig.~\ref{fig:GS}, are respectively 0.74(0.08), 0.70(0.08) and 0.52(0.13), showing that the ground states in the DSC regime belong mostly to the same parity chain due to the $Z_2$ symmetry of the QRM~\cite{casanova2010deep}. As the coupling ratio increases, the deviation from the ideal parity value +1 becomes more prominent due to imperfections of the adiabatic process and the motional heating arising from the occupation of larger Fock states.

By measuring the probability of recovering the initial state $\ket{\downarrow,0}$ after the ground state preparation and inverse process at time $t_{\rm Rev}$, we estimate a lower bound of the purity of the prepared ground state. The revival probabilities are 0.89(0.024), 0.87(0.027) and 0.79(0.03) for the three ratios $g/\omega_{\rm{m}}=1.2, 1.5$ and $2.0$, which give the lower bounds $0.79(0.042)$, $0.75(0.047)$ and $0.62(0.047)$, respectively. As shown in Fig. \ref{fig:GS}c, the reduced-spin purities, taking values $0.545(0.006)$, $0.514(0.003)$ and $0.505(0.002)$, are significantly smaller than the lower bounds of the total system, which prove the existence of entanglement within the prepared ground state at $t_{\rm Tar}$ (See Methods).

\section*{Spectrum}

\begin{figure*}[ht]
\includegraphics[width=0.8\textwidth]{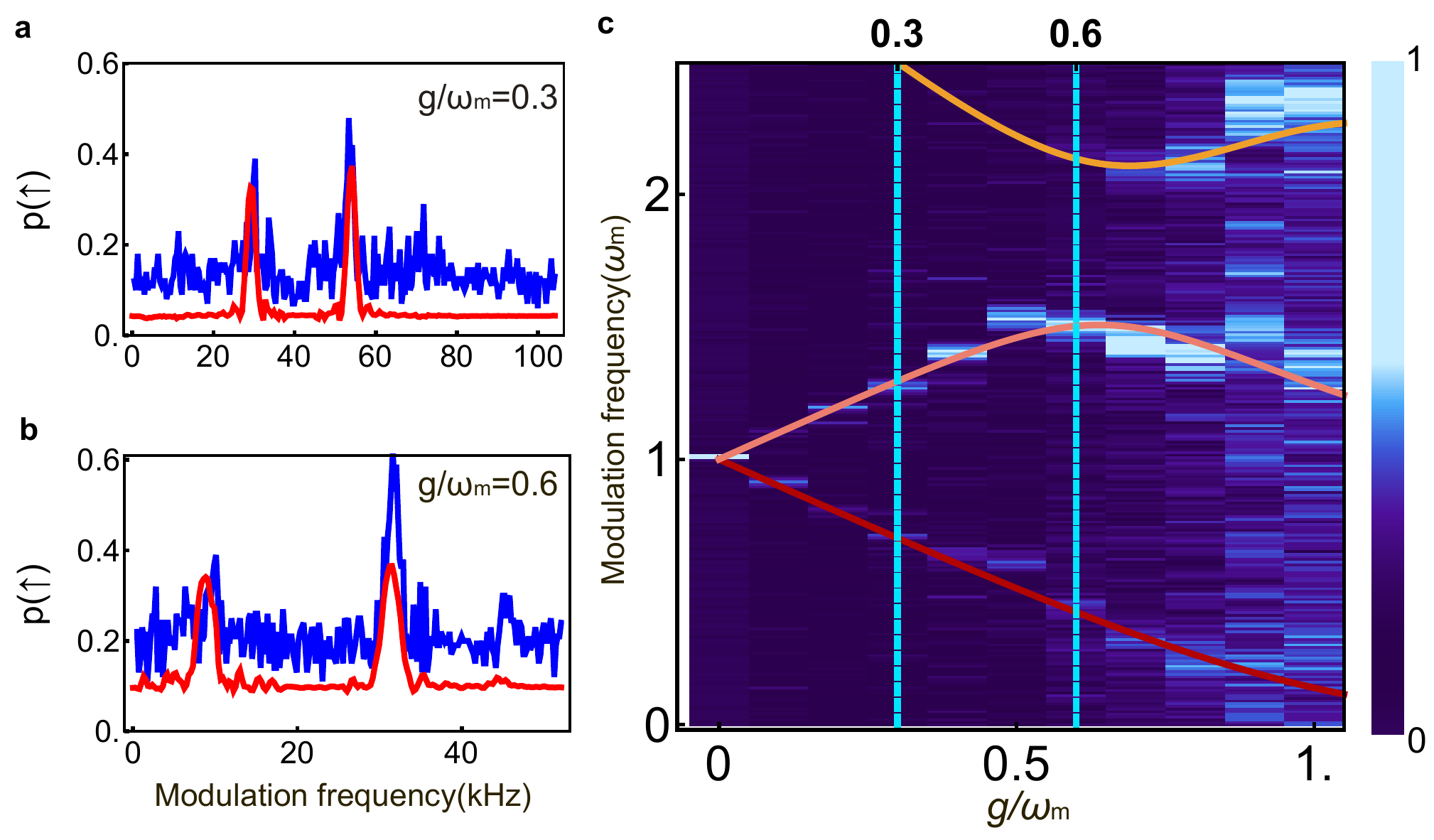}
\caption{ {\bf Spectrum of the QRM.} Panels (a) and (b) show, respectively for the regimes $\frac{g}{\omega_m }=0.3$ and $0.6$, the population of the excited state of the spin as a function of the modulation frequency of the probe driving. The red curve corresponds to numerical simulation results, while the blue curve shows the experimental data. Panel (c) shows the energy spectrum with the modulation frequency of the probe drive in the vertical axis rescaled by $\omega_{m}$. Note that the energy of the ground state (not plotted) is taken to be zero for all cases. The three continuous curves on top of the plot show the numerically computed energy spectrum of the states with parity opposite to the ground state.
\label{fig:Spectrum}}
\end{figure*}

The ground-state preparation can be extended to study the low-lying energy spectrum of the QRM by coherent spectroscopy~\cite{senko2014coherent}. In particular, we have measured the energy spectrum in the region $\frac{g}{\omega_m}\in[0,1]$. A $Z_2$ parity exists in the QRM model, which divides the Hilbert space in two, namely a subspace of parity $+1$ and other of parity $-1$~\cite{casanova2010deep}. Here, we focus on the energy splittings between the ground state and the first three excited states of opposite parity to the ground state~\cite{casanova2010deep}. For that, we have used a relatively weak modulated field as a probe on top of the simulation of the QRM, with the system initially in the ground state of the corresponding regime. We sweep the frequency of the probe pulse until we detect a transition, and we associate the frequency of the probe to the energy difference of the transition. To generate transitions between states of opposite parity, we use the probe pulse of the form
\begin{eqnarray}{\label{BreakParity}}
\hat{H}_{\rm{mod} } = \hat{H}_{\rm QRM}+
g_{\rm p} \sin(2\pi\nu_{\rm p} t)(\hat{\sigma}_{+}+\hat{\sigma}_{-}),
\end{eqnarray}
where $g_{\rm p}(\ll g)$ is the strength of the modulation field, and $\nu_{\rm p}$ is swept to find the resonant frequencies. In the region $\frac{g}{\omega_m }=0.$1 to $0.3$, $g_{\rm p}/g$ is 0.02, while the pulse duration is 350~$\mu s$. For the ratios $\frac{g}{\omega_{\rm m}} = 0.4$ to 1.0, the ratio $g_{\rm p}/g$ is 0.01, with a pulse duration of 450 $\us$. Population transfer is clearly seen when $\nu_p$ is resonant with the energy splittings as shown in Fig.~\ref{fig:Spectrum}.

\section*{Conclusion}
We have implemented the quantum simulation of all relevant coupling regimes of the QRM in a single trapped ion, obtaining direct evidence of the breakdown of the RWA. Historically, trapped ions have been always linked to the JCM physics, which has been enhanced here toward the more general QRM. In the DSC regime, we observe the phonon number wave-packets bounce back and forth as well as collapses and revivals of the initial state, confirming previous theoretical predictions. The adiabatic preparation of the ground state of the QRM in the DSC regime was produced for the first time in a quantum platform, and its reconstruction has enabled us to demonstrate the entanglement present in its ground state. As a direct application of this adiabatic method, we have been able to measure the energy splittings between states of different parity and recreate the energy spectrum of the QRM in the USC~regime. In conclusion, our work presents a detailed experimental exploration of the QRM in a wide range of physical regimes. Our experimental methods can be directly extended to the study of the phase transition in the QRM \cite{Hwang2015Quantum,Puebla2016Probing, hwang2017dissipative} or to the simulation of the Dicke model \cite{Dicke1954Coherence, Bastidas2012Nonequilibrium,Bakemeier2012Quantum} by considering the presence of more ions.

\section*{Methods}

\subsection{Calibration of the detuning of the blue and red sideband transitions}
For this experiment, since we fix the coupling strength $g$, the key ingredient for the simulation is to precisely set the detuning of the lasers.
For the ratio $\frac{g}{\omega_{\rm{m}} }=0.04$ case, since we set $\delta_{\rm{r}}=0$, and the $\delta_{\rm{b}}$ is much larger than the corresponding coupling strength, we obtain the resonance frequency of red-sideband transition with the detuning of the blue-sideband transition fixed at $\delta_{\rm{b}}=(2\pi)625\KHz$. For the USC/DSC regime, the coupling strength is comparable to the effective mode frequency, such that we need to carefully deal with the ac-Stark shift mainly introduced by the off-resonant excitation on the carrier transition. We measure the ac-Stark shift with a Ramsey experiment and calibrate the shift in the bichromatic pulse within $1\KHz$ accuracy. We further improve the frequency precision within 0.15 $\KHz$ range by giving the same detuning $\delta=(2\pi)10\KHz$ with different signs for both beams similarly to the scheme in Refs. \cite{lee2005phase,benhelm2008towards}.

\subsection{Phonon number state population distribution}
In Fig.~\ref{fig:GS}, we obtain the phonon number distribution. This is performed by driving the resonant blue-sideband transition $\ket{\downarrow,n} \leftarrow \ket{\uparrow,n+1}$ after the spin projective measurement and fitting the obtained spin population evolution with the formula \cite{meekhof1996generation,an2014experimental,lo2015spin}
\begin{eqnarray}
P_{\ket{\uparrow}}(t)=\frac{1}{2}\sum_{n}p(n)[1-\mbox{e}^{- \gamma t} \cos(\sqrt{n+1}\eta\Omega t)],
\label{RabiOS}
\end{eqnarray}
where $p(n)$ is the phonon number state population, $\gamma$ is the empirical decay constant, and $t$ is the pulse duration of the blue sideband.

\subsection{Verification of entanglement for the ground-state of the quantum Rabi model}
In the main text, we use ${\rm Tr}\left[\hat\rho_{\rm spin}^2\right]-P^2_{\rm Rev}<0$ to verify the existence of the entanglement between the spin and the phonon degrees of freedom. This can be understood as follows. First, we introduce the purity-based entanglement witness\cite{Islam2015Measuring,Horodecki2009Quantum} ${\mathcal W}$ for the target state $\hat\rho_{\rm Tar}$ at time $t_{\rm Tar}$, which is defined as
\begin{equation}\label{eq:entanglement_witness_sec2}
{\mathcal W}\left[\hat\rho_{\rm Tar}\right]\equiv{\rm Tr}\left[\hat\rho_{\rm spin}^2\right]-{\rm Tr}\left[\hat\rho_{\rm Tar}^2\right].
\end{equation}
It can be proved that ${\mathcal W}\left[\hat\rho\right]\geq0$ for arbitrary separable state. Thus ${\mathcal W}\left[\hat\rho\right]<0$ serves as the sufficient condition for the inseparability of $\hat\rho$.
However, the purity of the whole system ${\rm Tr}\left[\hat\rho_{\rm Tar}^2\right]$ requires full information of $\hat\rho_{\rm tot}$, which is quite demanding in our current experimental setup. Instead, we perform the disentangling operation, which is the time reversal of the adiabatic ground-state preparation in this particular case, and measure a single component
$P_{\rm Rev}\equiv Tr[|\downarrow,0\rangle \langle\downarrow,0|,\rho_{Rev}]$

of the spectral decomposition of the final state at time $t_{\rm Rev}$, which corresponds to the probability of the initial state . It's straightforward that $P_{\rm Rev}^2\leq{\rm Tr}\left[\hat\rho_{\rm Rev}\right]$. With reasonable assumption that the evolution performed in this experiment never increases purity, we can push the inequality to an earlier time. Then we have the following inequality
\begin{equation}\label{eq:inequlity_sec2}
{\rm Tr}\left[\hat\rho_{\rm Tar}^2\right]\geq{\rm Tr}\left[\hat\rho_{\rm Rev}^2\right]\geq P_{\rm Rev}^2.
\end{equation}
In other word, $P_{\rm Rev}$ serves a lower bound for ${\rm Tr}\left[\hat\rho_{\rm Tar}^2\right]$. Putting Eqs.~(\ref{eq:entanglement_witness_sec2}) and (\ref{eq:inequlity_sec2}) together, we have
\begin{equation}
{\mathcal W}\left[\hat \rho_{\rm Tar}\right]\leq{\rm Tr}\left[\hat\rho_{\rm spin}^2\right]-P^2_{\rm Rev}.
\end{equation}

\section*{Acknowledgements}
We thank Xiao Yuan, Xiongfeng Ma, Hyunchul Nha, Jiyong Park, Jae-hak Lee, and M. S. Kim for useful discussions on the entanglement verification of the ground state. This work was supported by the National Key Research and Development Program of China under Grants No. 2016YFA0301900 and No. 2016YFA0301901 and the National Natural Science Foundation of China Grants No. 11374178, No. 11574002, and No. 11504197, MINECO/FEDER FIS2015-69983-P, Ram\'on y Cajal Grant RYC-2012-11391, and Basque Government IT986-16.


\begin{thebibliography}{10}
\expandafter\ifx\csname url\endcsname\relax
  \def\url#1{\texttt{#1}}\fi
\expandafter\ifx\csname urlprefix\endcsname\relax\def\urlprefix{URL }\fi
\providecommand{\bibinfo}[2]{#2}
\providecommand{\eprint}[2][]{\url{#2}}

\bibitem{rabi1936process}
\bibinfo{author}{Rabi, I.}
\newblock \bibinfo{title}{On the process of space quantization}.
\newblock \emph{\bibinfo{journal}{Physical Review}}
  \textbf{\bibinfo{volume}{49}}, \bibinfo{pages}{324} (\bibinfo{year}{1936}).

\bibitem{frisch1933einstellung}
\bibinfo{author}{Frisch, R.} \& \bibinfo{author}{Segre, E.}
\newblock \bibinfo{title}{{\"U}ber die einstellung der richtungsquantelung.
  ii}.
\newblock \emph{\bibinfo{journal}{Zeitschrift f{\"u}r Physik A Hadrons and
  Nuclei}} \textbf{\bibinfo{volume}{80}}, \bibinfo{pages}{610--616}
  (\bibinfo{year}{1933}).

\bibitem{Jaynes63}
\bibinfo{author}{Jaynes, E.~T.} \& \bibinfo{author}{Cummings, F.~W.}
\newblock \bibinfo{title}{Comparison of quantum and semiclassical radiation
  theories with application to the beam maser}.
\newblock \emph{\bibinfo{journal}{Proceedings of the IEEE}}
  \textbf{\bibinfo{volume}{51}}, \bibinfo{pages}{89--109}
  (\bibinfo{year}{1963}).

\bibitem{miller2005trapped}
\bibinfo{author}{Miller, R.} \emph{et~al.}
\newblock \bibinfo{title}{Trapped atoms in cavity qed: coupling quantized light
  and matter}.
\newblock \emph{\bibinfo{journal}{Journal of Physics B: Atomic, Molecular and
  Optical Physics}} \textbf{\bibinfo{volume}{38}}, \bibinfo{pages}{S551}
  (\bibinfo{year}{2005}).

\bibitem{walther2006cavity}
\bibinfo{author}{Walther, H.}, \bibinfo{author}{Varcoe, B.~T.},
  \bibinfo{author}{Englert, B.-G.} \& \bibinfo{author}{Becker, T.}
\newblock \bibinfo{title}{Cavity quantum electrodynamics}.
\newblock \emph{\bibinfo{journal}{Reports on Progress in Physics}}
  \textbf{\bibinfo{volume}{69}}, \bibinfo{pages}{1325} (\bibinfo{year}{2006}).

\bibitem{Haroche}
\bibinfo{author}{Raimond, J.-M.} \& \bibinfo{author}{Haroche, S.}
\newblock \emph{\bibinfo{title}{Exploring the quantum}}
  (\bibinfo{publisher}{Oxford University Press, Oxford}, \bibinfo{year}{2006}).

\bibitem{Ritsch2013Cold}
\bibinfo{author}{Ritsch, H.}, \bibinfo{author}{Domokos, P.},
  \bibinfo{author}{Brennecke, F.} \& \bibinfo{author}{Esslinger, T.}
\newblock \bibinfo{title}{Cold atoms in cavity-generated dynamical optical
  potentials}.
\newblock \emph{\bibinfo{journal}{Rev. Mod. Phys.}}
  \textbf{\bibinfo{volume}{85}}, \bibinfo{pages}{553--601}
  (\bibinfo{year}{2013}).

\bibitem{Hanson2006Spins}
\bibinfo{author}{Hanson, R.}, \bibinfo{author}{Kouwenhoven, L.~P.},
  \bibinfo{author}{Petta, J.~R.}, \bibinfo{author}{Tarucha, S.} \&
  \bibinfo{author}{Vandersypen, L. M.~K.}
\newblock \bibinfo{title}{Spins in few-electron quantum dots}.
\newblock \emph{\bibinfo{journal}{Rev. Mod. Phys.}}
  \textbf{\bibinfo{volume}{79}}, \bibinfo{pages}{1217--1265}
  (\bibinfo{year}{2006}).

\bibitem{wallraff2004strong}
\bibinfo{author}{Wallraff, A.}, \bibinfo{author}{Schuster, D.~I.},
  \bibinfo{author}{Blais, A.}, \bibinfo{author}{Frunzio, L.} \emph{et~al.}
\newblock \bibinfo{title}{Strong coupling of a single photon to a
  superconducting qubit using circuit quantum electrodynamics}.
\newblock \emph{\bibinfo{journal}{Nature}} \textbf{\bibinfo{volume}{431}},
  \bibinfo{pages}{162} (\bibinfo{year}{2004}).

\bibitem{devoret2013superconducting}
\bibinfo{author}{Devoret, M.~H.} \& \bibinfo{author}{Schoelkopf, R.~J.}
\newblock \bibinfo{title}{Superconducting circuits for quantum information: an
  outlook}.
\newblock \emph{\bibinfo{journal}{Science}} \textbf{\bibinfo{volume}{339}},
  \bibinfo{pages}{1169--1174} (\bibinfo{year}{2013}).

\bibitem{meekhof1996generation}
\bibinfo{author}{Meekhof, D.}, \bibinfo{author}{Monroe, C.},
  \bibinfo{author}{King, B.}, \bibinfo{author}{Itano, W.~M.} \&
  \bibinfo{author}{Wineland, D.~J.}
\newblock \bibinfo{title}{Generation of nonclassical motional states of a
  trapped atom}.
\newblock \emph{\bibinfo{journal}{Phys. Rev. Lett.}}
  \textbf{\bibinfo{volume}{76}}, \bibinfo{pages}{1796} (\bibinfo{year}{1996}).

\bibitem{Leibfried03}
\bibinfo{author}{Leibfried, D.}, \bibinfo{author}{Blatt, R.},
  \bibinfo{author}{Monroe, C.} \& \bibinfo{author}{Wineland, D.~J.}
\newblock \bibinfo{title}{Quantum dynamics of single trapped ions}.
\newblock \emph{\bibinfo{journal}{Rev. Mod. Phys.}}
  \textbf{\bibinfo{volume}{75}}, \bibinfo{pages}{281--324}
  (\bibinfo{year}{2003}).

\bibitem{Haffner08}
\bibinfo{author}{H{\"a}ffner, H.}, \bibinfo{author}{Roos, C.~F.} \&
  \bibinfo{author}{Blatt, R.}
\newblock \bibinfo{title}{Quantum computing with trapped ions}.
\newblock \emph{\bibinfo{journal}{Phys. Rep.}} \textbf{\bibinfo{volume}{469}},
  \bibinfo{pages}{155--203} (\bibinfo{year}{2008}).

\bibitem{lv2017reconstruction}
\bibinfo{author}{Lv, D.} \emph{et~al.}
\newblock \bibinfo{title}{Reconstruction of the jaynes-cummings field state of
  ionic motion in a harmonic trap}.
\newblock \emph{\bibinfo{journal}{Phys. Rev. A}} \textbf{\bibinfo{volume}{95}},
  \bibinfo{pages}{043813} (\bibinfo{year}{2017}).

\bibitem{niemczyk2010circuit}
\bibinfo{author}{Niemczyk, T.} \emph{et~al.}
\newblock \bibinfo{title}{Circuit quantum electrodynamics in the
  ultrastrong-coupling regime}.
\newblock \emph{\bibinfo{journal}{Nature Phys.}} \textbf{\bibinfo{volume}{6}},
  \bibinfo{pages}{772--776} (\bibinfo{year}{2010}).

\bibitem{Braum2016Analog}
\bibinfo{author}{Braumüller, J.} \emph{et~al.}
\newblock \bibinfo{title}{Analog quantum simulation of the rabi model in the
  ultra-strong coupling regime.}
\newblock \emph{\bibinfo{journal}{Nature Commun.}} \textbf{\bibinfo{volume}{8}}
  (\bibinfo{year}{2016}).

\bibitem{forn2017ultrastrong}
\bibinfo{author}{Forn-D{\'\i}az, P.} \emph{et~al.}
\newblock \bibinfo{title}{Ultrastrong coupling of a single artificial atom to
  an electromagnetic continuum in the nonperturbative regime}.
\newblock \emph{\bibinfo{journal}{Nature Phys.}} \textbf{\bibinfo{volume}{13}},
  \bibinfo{pages}{39--43} (\bibinfo{year}{2017}).

\bibitem{yoshihara2017superconducting}
\bibinfo{author}{Yoshihara, F.} \emph{et~al.}
\newblock \bibinfo{title}{Superconducting qubit-oscillator circuit beyond the
  ultrastrong-coupling regime}.
\newblock \emph{\bibinfo{journal}{Nature Phys.}} \textbf{\bibinfo{volume}{13}},
  \bibinfo{pages}{44--47} (\bibinfo{year}{2017}).

\bibitem{Rossatto2016Spectral}
\bibinfo{author}{Rossatto, D.~Z.}, \bibinfo{author}{Villasbôas, C.~J.},
  \bibinfo{author}{Sanz, M.} \& \bibinfo{author}{Solano, E.}
\newblock \bibinfo{title}{Spectral classification of coupling regimes in the
  quantum rabi model}.
\newblock \emph{\bibinfo{journal}{Phys. Rev. A}} \textbf{\bibinfo{volume}{96}},
  \bibinfo{pages}{013849} (\bibinfo{year}{2016}).

\bibitem{casanova2010deep}
\bibinfo{author}{Casanova, J.}, \bibinfo{author}{Romero, G.},
  \bibinfo{author}{Lizuain, I.}, \bibinfo{author}{Garc{\'\i}a-Ripoll, J.~J.} \&
  \bibinfo{author}{Solano, E.}
\newblock \bibinfo{title}{Deep strong coupling regime of the jaynes-cummings
  model}.
\newblock \emph{\bibinfo{journal}{Phys. Rev. Lett.}}
  \textbf{\bibinfo{volume}{105}}, \bibinfo{pages}{263603}
  (\bibinfo{year}{2010}).

\bibitem{romero2012ultrafast}
\bibinfo{author}{Romero, G.}, \bibinfo{author}{Ballester, D.},
  \bibinfo{author}{Wang, Y.}, \bibinfo{author}{Scarani, V.} \&
  \bibinfo{author}{Solano, E.}
\newblock \bibinfo{title}{Ultrafast quantum gates in circuit qed}.
\newblock \emph{\bibinfo{journal}{Phys. Rev. Lett.}}
  \textbf{\bibinfo{volume}{108}}, \bibinfo{pages}{120501}
  (\bibinfo{year}{2012}).

\bibitem{braak2011integrability}
\bibinfo{author}{Braak, D.}
\newblock \bibinfo{title}{Integrability of the rabi model}.
\newblock \emph{\bibinfo{journal}{Phys. Rev. Lett.}}
  \textbf{\bibinfo{volume}{107}}, \bibinfo{pages}{100401}
  (\bibinfo{year}{2011}).

\bibitem{crespi2012photonic}
\bibinfo{author}{Crespi, A.}, \bibinfo{author}{Longhi, S.} \&
  \bibinfo{author}{Osellame, R.}
\newblock \bibinfo{title}{Photonic realization of the quantum rabi model}.
\newblock \emph{\bibinfo{journal}{Phys. Rev. Lett.}}
  \textbf{\bibinfo{volume}{108}}, \bibinfo{pages}{163601}
  (\bibinfo{year}{2012}).

\bibitem{langford2016experimentally}
\bibinfo{author}{Langford, N.} \emph{et~al.}
\newblock \bibinfo{title}{Experimentally simulating the dynamics of quantum
  light and matter at ultrastrong coupling}.
\newblock \emph{\bibinfo{journal}{arXiv preprint arXiv:1610.10065}}
  (\bibinfo{year}{2016}).

\bibitem{pedernales2015quantum}
\bibinfo{author}{Pedernales, J.} \emph{et~al.}
\newblock \bibinfo{title}{Quantum rabi model with trapped ions}.
\newblock \emph{\bibinfo{journal}{Scientific Reports}}
  \textbf{\bibinfo{volume}{5}}, \bibinfo{pages}{15472} (\bibinfo{year}{2015}).

\bibitem{an2014experimental}
\bibinfo{author}{An, S.} \emph{et~al.}
\newblock \bibinfo{title}{Experimental test of the quantum jarzynski equality
  with a trapped-ion system}.
\newblock \emph{\bibinfo{journal}{Nature Phys.}}  (\bibinfo{year}{2014}).

\bibitem{sidebandcooling}
\bibinfo{author}{Monroe, C.} \emph{et~al.}
\newblock \bibinfo{title}{Resolved-sideband raman cooling of a bound atom to
  the 3d zero-point energy}.
\newblock \emph{\bibinfo{journal}{Phys. Rev. Lett.}}
  \textbf{\bibinfo{volume}{75}}, \bibinfo{pages}{4011} (\bibinfo{year}{1995}).

\bibitem{S2000Entanglement}
\bibinfo{author}{Sørensen, A.} \& \bibinfo{author}{Mølmer, K.}
\newblock \bibinfo{title}{Entanglement and quantum computation with ions in
  thermal motion}.
\newblock \emph{\bibinfo{journal}{Phys. Rev. A}} \textbf{\bibinfo{volume}{62}},
  \bibinfo{pages}{117--134} (\bibinfo{year}{2000}).

\bibitem{Haljan2005Spin}
\bibinfo{author}{Haljan, P.~C.}, \bibinfo{author}{Brickman, K.~A.},
  \bibinfo{author}{Deslauriers, L.}, \bibinfo{author}{Lee, P.~J.} \&
  \bibinfo{author}{Monroe, C.}
\newblock \bibinfo{title}{Spin-dependent forces on trapped ions for
  phase-stable quantum gates and entangled states of spin and motion.}
\newblock \emph{\bibinfo{journal}{Phys. Rev. Lett.}}
  \textbf{\bibinfo{volume}{94}}, \bibinfo{pages}{153602}
  (\bibinfo{year}{2005}).

\bibitem{Nielson2000Quantum}
\bibinfo{author}{Nielson, M.~A.} \& \bibinfo{author}{Chuang, I.~L.}
\newblock \emph{\bibinfo{title}{Quantum Computation and Quantum Information}}
  (\bibinfo{publisher}{Cambridge University Press,}, \bibinfo{year}{2000}).

\bibitem{senko2014coherent}
\bibinfo{author}{Senko, C.} \emph{et~al.}
\newblock \bibinfo{title}{Coherent imaging spectroscopy of a quantum many-body
  spin system}.
\newblock \emph{\bibinfo{journal}{Science}} \textbf{\bibinfo{volume}{345}},
  \bibinfo{pages}{430--433} (\bibinfo{year}{2014}).

\bibitem{Hwang2015Quantum}
\bibinfo{author}{Hwang, M.~J.}, \bibinfo{author}{Puebla, R.} \&
  \bibinfo{author}{Plenio, M.~B.}
\newblock \bibinfo{title}{Quantum phase transition and universal dynamics in
  the rabi model.}
\newblock \emph{\bibinfo{journal}{Phys. Rev. Lett.}}
  \textbf{\bibinfo{volume}{115}} (\bibinfo{year}{2015}).

\bibitem{Puebla2016Probing}
\bibinfo{author}{Puebla, R.}, \bibinfo{author}{Hwang, M.~J.},
  \bibinfo{author}{Casanova, J.} \& \bibinfo{author}{Plenio, M.~B.}
\newblock \bibinfo{title}{Probing the dynamics of superradiant quantum phase
  transition in a single trapped-ion}.
\newblock \emph{\bibinfo{journal}{Phys. Rev. Lett.}}
  \textbf{\bibinfo{volume}{118}} (\bibinfo{year}{2016}).

\bibitem{hwang2017dissipative}
\bibinfo{author}{Hwang, M.-J.}, \bibinfo{author}{Rabl, P.} \&
  \bibinfo{author}{Plenio, M.~B.}
\newblock \bibinfo{title}{Dissipative phase transition in the open quantum rabi
  model}.
\newblock \emph{\bibinfo{journal}{arXiv preprint arXiv:1708.08175}}
  (\bibinfo{year}{2017}).

\bibitem{Dicke1954Coherence}
\bibinfo{author}{Dicke, R.~H.}
\newblock \bibinfo{title}{Coherence in spontaneous radiation processes}.
\newblock \emph{\bibinfo{journal}{Physical Review}}
  \textbf{\bibinfo{volume}{93}}, \bibinfo{pages}{99--110}
  (\bibinfo{year}{1954}).

\bibitem{Bastidas2012Nonequilibrium}
\bibinfo{author}{Bastidas, V.~M.}, \bibinfo{author}{Emary, C.},
  \bibinfo{author}{Regler, B.} \& \bibinfo{author}{Brandes, T.}
\newblock \bibinfo{title}{Nonequilibrium quantum phase transitions in the dicke
  model.}
\newblock \emph{\bibinfo{journal}{Phys. Rev. Lett.}}
  \textbf{\bibinfo{volume}{108}}, \bibinfo{pages}{043003}
  (\bibinfo{year}{2012}).

\bibitem{Bakemeier2012Quantum}
\bibinfo{author}{Bakemeier, L.}, \bibinfo{author}{Alvermann, A.} \&
  \bibinfo{author}{Fehske, H.}
\newblock \bibinfo{title}{Quantum phase transition in the dicke model with
  critical and noncritical entanglement}.
\newblock \emph{\bibinfo{journal}{Phys. Rev. A}} \textbf{\bibinfo{volume}{85}},
  \bibinfo{pages}{1073--1079} (\bibinfo{year}{2012}).

\bibitem{lee2005phase}
\bibinfo{author}{Lee, P.} \emph{et~al.}
\newblock \bibinfo{title}{Phase control of trapped ion quantum gates}.
\newblock \emph{\bibinfo{journal}{Journal of Optics B: Quantum and
  Semiclassical Optics}} \textbf{\bibinfo{volume}{7}}, \bibinfo{pages}{S371}
  (\bibinfo{year}{2005}).

\bibitem{benhelm2008towards}
\bibinfo{author}{Benhelm, J.}, \bibinfo{author}{Kirchmair, G.},
  \bibinfo{author}{Roos, C.~F.} \& \bibinfo{author}{Blatt, R.}
\newblock \bibinfo{title}{Towards fault-tolerant quantum computing with trapped
  ions}.
\newblock \emph{\bibinfo{journal}{Nature Phys.}} \textbf{\bibinfo{volume}{4}},
  \bibinfo{pages}{463--466} (\bibinfo{year}{2008}).

\bibitem{lo2015spin}
\bibinfo{author}{Lo, H.-Y.} \emph{et~al.}
\newblock \bibinfo{title}{Spin-motion entanglement and state diagnosis with
  squeezed oscillator wavepackets}.
\newblock \emph{\bibinfo{journal}{Nature}} \textbf{\bibinfo{volume}{521}},
  \bibinfo{pages}{336--339} (\bibinfo{year}{2015}).

\bibitem{Islam2015Measuring}
\bibinfo{author}{Islam, R.} \emph{et~al.}
\newblock \bibinfo{title}{Measuring entanglement entropy in a quantum many-body
  system}.
\newblock \emph{\bibinfo{journal}{Nature}} \textbf{\bibinfo{volume}{528}},
  \bibinfo{pages}{77--83} (\bibinfo{year}{2015}).

\bibitem{Horodecki2009Quantum}
\bibinfo{author}{Horodecki, R.}, \bibinfo{author}{Horodecki, P.},
  \bibinfo{author}{Horodecki, M.} \& \bibinfo{author}{Horodecki, K.}
\newblock \bibinfo{title}{Quantum entanglement}.
\newblock \emph{\bibinfo{journal}{Rev. Mod. Phys.}}
  \textbf{\bibinfo{volume}{81}}, \bibinfo{pages}{865--942}
  (\bibinfo{year}{2009}).

\end{thebibliography}

\end{document}